\documentclass{article}
\usepackage{amsfonts,amsmath,amssymb,amscd,euscript}
\usepackage[english]{babel}

\newtheorem{pr}{Lemma}
\newtheorem{Th}{Theorem}
\def\al{\alpha}
\newcommand{\e}{\varepsilon}
\newcommand{\eu}{\EuScript}
\renewcommand{\l}{\label}
\newcommand{\n}{\nabla}

\def\t{\tilde}

\def\be{\begin{equation}} \def\ee{\end{equation}}
\def\bg{\begin{gather}} \def\eg{\end{gather}}
\def\ba{\begin{align}} \def\ea{\end{align}}
\def\bea{\begin{eqnarray}} \def\eea{\end{eqnarray}}
\newcommand{\bt}{\begin{Th}} \newcommand{\et}{\end{Th}}
\newcommand{\bp}{\begin{pr}} \newcommand{\ep}{\end{pr}}

\begin{document}

\centerline{\large \bf Effective connections and fields}
\bigskip
\centerline{\large \bf associated with shear-free null congruences}
\bigskip
\centerline{Vladimir V.~Kassandrov and Vladimir N.~Trishin
 \footnote{Department of General Physics, Peoples' Friendship University of Russia,
        Ordjonikidze 3, 117419, Moscow, Russia. \\
        E-mail:~vkassan@sci.pfu.edu.ru}
}
\bigskip
\begin{abstract}

A special subclass of shear-free null congruences (SFC) is studied, with tangent vector
field being a repeated principal null direction of the Weyl tensor. We demonstrate that
this field is parallel with respect to an effective affine connection which contains the
Weyl nonmetricity and the skew symmetric torsion. On the other hand, a Maxwell-like field
can be directly associated with any special SFC, and the electric charge for bounded
singularities of this field turns to be ``self-quantized''. Two invariant differential
operators are introduced which can be thought of as spinor analogues of the Beltrami
operators and both nullify the principal spinor of any special SFC.
\end{abstract}
\bigskip


\bigskip

\section{Introduction}

Among the congruences of null geodesics (of rectilinear light rays in flat Minkowski
space) the congruences with zero shear (shear-free congruences, SFC) are certainly
distinguished. Recall that {\it expansion, twist} and {\it shear} are three {\it optical
invariants}~\cite{PR86} which can be associated with every geodesic congruence and
describe the deformations of infinitesimal orthogonal sections along the rays.
Particularly, light rays emitted by an arbitrary moving point source form a simplest SFC
(with zero twist)~\cite{K69}. Complexification of this construction leads to twisting SFC
including the famous Kerr congruence.

In flat (precisely, in conformally flat) space, SFC are closely related to
{\it twistor} geometry, and all analytical SFC can be obtained algebraically
via the twistor construction and the {\it Kerr theorem}~\cite{PR86}. These
congruences represent also one of the two classes of solutions to complex
eikonal equation~\cite{K2002}. On the other hand, SFC naturally arise in the
framework of noncommutative analysis (over the algebra of biquaternions),
and their defining equation turns to be equivalent to the (nonlinear)
generalization of the Cauchy-Riemann differentiability conditions~\cite{acta,KR2000}.
On these grounds, an unified algebraic field theory (nonlinear,
over-determined, non-Lagrangian) has been developed in our
works~\cite{K95,trish,KR2000,K2002} in which the approach has been called
{\it algebrodynamics}.

Indeed, SFC in flat space manifest also numerous connections with
equations for massless fields (Penrose transform and Ward construction
~\cite{Ward} for self-dual fields among them ) but the interlinks are not
direct. The first and the most known result which relates the SFC and the solutions
to massless field equations (to Maxwell equations in particular)
is the Robinson's theorem~\cite{Rob61}.
It asserts that there always exists an affine parameter along the congruence such
that the self-dual 2-form  $F_{\mu\nu}:=\e_{A'B'}\xi_A\xi_B$ defined via the
the principal spinor $\xi_A$ of the SFC is {\it exact} and, therefore,
satisfies the homogeneous Maxwell equations. However, electromagnetic
field obtained in this way is null, and there is no straightforward
generalization of the Robinson theorem to general (non-null) fields. We mention
here only the works~\cite{BCK97,CK79} where SFC were applied to construct
the Hertz potential of an electromagnetic field on  Riemannian space-times.

On the other hand, in the framework of algebrodynamics it was shown that every
SFC induces an effective affine connection with a Weyl nonmetricity and a
skew symmetric torsion~\cite{K95}. Defining equations for
SFC become then just the conditions for principal spinor of SFC to be
{\it covariantly constant} (parallel) with respect to this Weyl-Cartan
connection.

In analogy with the unified Weyl theory, for any SFC one can define, therefore,
a {\it gauge field} with potentials represented by the Weyl 4-vector. From
the integrability conditions it follows then that correspondent field strengths
should satisfy homogeneous Maxwell equations. Further on we refer to this field
as the {\it gauge field of the congruence} (GFC). Besides, the potentials of an
$SL(2,\mathbb{C})$ matrix gauge field can be defined via the components of the same
Weyl vector, and by virtue of integrability conditions {\it it obeys
the Yang-Mills type equations}~\cite{K95}.  It is known also ~\cite{KW79} (see
also~\cite{KR2000}) that the component of projective spinor defining tangent
vector field of a SFC necessarily satisfies both linear wave and nonlinear
eikonal equations.

Note that Maxwell field (GFC) which arises in the above procedure has a lot of remarkable
properties which are not inherent to fields introduced by Robinson and Penrose.
Specifically, this field is really a gauge one possessing a residual ``weak''
gauge group closely related to transformations in the projective twistor space~\cite{KR2000}.
Moreover, the strengths and the singular loci of the GFC admit explicit representation
in twistor variables~\cite{KR2000,K2002} and can be obtained in a completely
algebraic way. Finally, effective {\it electric charge} of bounded singularities
of the GFC turns to be {\it self-quantized}, i.e. integer multiple to a minimal
``elementary'' one~\cite{K95, prot, K2003}.

It is also well known that an effective Riemannian metric of the Kerr-Schild type can be
canonically associated with every SFC in Minkowski space. Under this particular
deformation of space-time geometry the congruence preserves all its properties, i.e.
remains null, geodetic and shear-free. Singularities of curvature of the deformed metrics
are completely determined by the congruence and correspond to the locus of {\it branching
points} of the latter. It is especially interesting that curvature singularities and
strength singularities of the above discussed GFC completely {\it coincide} in space and
time and define, therefore, a unique {\it particle-like} object with nontrivial time
evolution. Note also that in many cases the induced metric can be choosed to satisfy
vacuum or electrovacuum Einstein equations: for instance, the Kerr and the Kerr-Newman
metrics can be obtained in such a way.

In general relativity significance of SFC is also justified by the Goldberg-Sachs
theorem~\cite{GoldSach}: for any Einstein space the Weyl conformal tensor is
algebraically special iff the manifold admits a SFC which in this case defines
one of its {\it repeated} principal null direction (PND). Particularly, it follows that
in any {\it vacuum} space-time there are at most 2 independent SFC (for the
space of Petrov type D) or a single SFC in spaces of type II, III or N.
The generalization of this theorem is known~\cite{Somm76}.
Generically, if a manifold is not conformally Einstein (and, therefore, not
conformally flat) the number of distinct SFC {\it can't exceed four} because
every SFC necessarily defines a PND of the Weyl tensor~
\footnote{In fact, only
an example of a space with {\it three} independent SFC was presented in~\cite{Nurowski}}.
Recall for comparison that in a conformally flat space there exists an
{\it infinity} of distinct SFC described by the Kerr theorem.

In the paper, we restrict ourselves by a special subclass of SFC which define a repeated
PND of the Weyl spinor and can, thus, exist only on an algebraically special Riemannian
space-time. We prove that, as it takes place for the flat case, any such SFC defines a
vector field parallel with respect to an effective Weyl-Cartan connection. Further on we
consider the fields which can be associated with a special SFC and which are, in fact,
inherited from those defined for rectilinear SFC on the flat background and described
above. In particular, we show that the GFC correspondent to a special SFC on an
asymptotically flat space preserves its Coulomb-like structure and the property of
self-quantization of effective electric charge. As as example, the situation for radial
(special) SFC on the Schwarzschild  background is analyzed. Finally, we adapt the wave
and the eikonal Beltrami operators to the 2-spinor fields and demonstrate that they both
nullify the principal spinor of any special SFC. This property also can be looked at as a
generalization of similar property of SFC on the Minkowski space-time.

The notation used in the paper corresponds to the standard one~\cite{PR86}.

\section{Special shear - free null congruences and related Weyl - Cartan
connections}

Let $l^\mu$ be a null vector field tangent to the rays of a null congruence,
and $\xi_A$ be a related {\it principal} spinor field of the congruence.
In abstract index notation~\cite{PR86} the correspondence is
settled by $l^\mu=\xi^A\xi^{A'}$. Condition for the congruence to be
shear-free (and, therefore, geodetic) is then
\be \l{SFC}
\xi^A\xi^B\nabla_{AA'}\xi_B=0 ,
\ee
where $\nabla_{AA'}$ stands for the spinor derivative with respect to the Levi-Civita
connection. Making use of the spinor algebra, it is easy to check that SFC defining
condition (\ref{SFC}) is equivalent to one of the following two forms:
\be \l{SFC2}
\nabla_{A'(A}\xi_{B)}=\phi_{A'(A}\xi_{B)}
\ee
or
\be \l{SFC3}
\xi^A\nabla_{AA'}\xi_B=\eta_{A'}\xi_B
\ee
where an auxiliary complex vector field $\phi_{A'A}$ and a 2-spinor field
$\eta_{A'}$ are introduced. It follows then that covariant derivative of the
principal spinor can be always represented in the form
\be \l{cd}
\nabla_{AA'}\xi_B=\phi_{BA'}\xi_A+\varepsilon_{AB}\eta_{A'}
\ee

Eq.(\ref{SFC}) and therefore Eqs.(\ref{SFC2}),(\ref{SFC3}) are invariant under
scalings of the principal spinor of the form
\bg\l{scale}
\xi_A\mapsto\hat\xi_A=\alpha\xi_A \\
\phi_\mu\mapsto\hat\phi_\mu=\phi_\mu+\n_\mu \ln \alpha,~~~
\eta_{A'}\mapsto\hat\eta_{A'}=\al(\eta_{A'}+\xi^A\n_{AA'}\ln\al)
\end{gather}
with $\alpha(x)$ being an arbitrary (differentiable) complex function. In
flat case it can be fixed so that the spinor $\eta(x)$
identically vanishes and SFC condition (\ref{SFC}) reduces to the
form~\cite{KR2000}
\be \l{red}
\xi^A\nabla_{AA'}\xi_B=0~~~\Leftrightarrow~~~\nabla_{AA'}\xi_B=\phi_{BA'}\xi_A
\ee
Properties of the {\it gauge field of the congruence} (GFC) represented by
4-potentials $\phi_{BA'}$ in Eq.(\ref{red}) which in the flat case satisfies
homogeneous Maxwell equations and for which the value of effective electric
charge is necessarily self-quantized have been presented in the introduction.
Further on we return to study its properties on a curved Riemannian background.

Consider now the reduced form of SFC condition (\ref{red}). Differentiating it
and commuting the spinor covariant derivatives in the l.h.s. we get
\be\l{integred}
\Psi_{ABCD}\xi^B\xi^C\xi^D=0,
\ee
where $\Psi_{ABCD}=\Psi_{(ABCD)}$ is the Weyl spinor of conformal curvature.
Thus, apart of the conformally flat case, reduction of the SFC equation to the
form (\ref{red}) can be made only for {\it algebraically special} spaces
and only when the particular SFC defines a {\it repeated} principal null
direction (RPND) of the Weyl conformal tensor. Recall by this that integrability
condition for generic form of SFC equation (\ref{SFC}) leads to a weaker
condition  $\Psi_{ABCD}\xi^A\xi^B\xi^C\xi^D=0$ which does not
restrict the curvature implying only for the principal SFC spinor to
represent one of the four null directions of the Weyl tensor (not necessarily
repeated).

Conversely, let now an algebraically special space is equipped with a SFC
defining a RPND of the Weyl tensor. Can its equation be reduced to the form
(\ref{red}), i.e. the spinor $\eta_{A'}$  be turned into zero? To ensure
this, a spinor $\eta$ being given, the scaling parameter $\alpha(x)$ must be
found which obeys the equation $\xi^A\n_{AA'}\ln\al=-\eta_{A'}$. Calvulating
the commutator of the derivatives and taking into account that the spinor
$\eta_{A'}$ is related to $\xi_A$ by (\ref{SFC3}) we obtain the following
necessary condition for the spinor $\hat\eta_{A'}$ to vanish:
\be \l{somm}
\xi^A\n_{AA'}\eta^{A'}=0.
\ee
For analytic congruences this is also a sufficient one \cite{Somm76}. Using
then Eq.(\ref{SFC3}) and taking the contraction $(\xi^A{\nabla_A}^{C'})
(\xi^C\nabla_{CC'})\xi_B$ in its l.h.s. we obtain that condition (\ref{somm})
is equivalent to the old one $\Psi_{ABCD}\xi^B\xi^C\xi^D=0$ so that
necessary condition (\ref{integred}) is also a sufficient one.
Note that if the metric is conformally flat the IC are satisfied identically.
So we have
\begin{pr}
The SFC equation (\ref{SFC}) can be represented in the form (\ref{red})
iff the SFC spinor is a RPND of the Weyl tensor or the space is conformally flat.
\end{pr}



Further on SFC which admit a reduced representation (\ref{red}) and define
a RPND of the Weyl tensor will be refered to as {\it special} SFC (SSFC).

Let us notice now that another evident form of representation of SSFC equation
(\ref{red}) is the condition for its principal spinor $\xi_A$ to be
{\it covariantly constant} ({\it parallel}),
\be \l{parallel}
\tilde\n_{AA'} \xi_B = 0 ,
\ee
with respect to an effective spinor connection represented by
\be \l{spweylc}
\tilde\n_{AA'}\xi_B\equiv\n_{AA'}\xi_B - {\Gamma_{AA'B}}^C\xi_C, \qquad
{\Gamma_{AA'B}}^C=\phi_{BA'}{\e_A}^C
\ee.
Thus Eq.(\ref{parallel}) is a spinor image of the condition for tangent vector of
the SSFC $l_\mu$ to be parallel with respect to affine connection of the form
\be\l{realweylc}
\tilde\n_\mu l_\al = \n_\mu l_\al - 2({\delta_\mu}^\beta a_\alpha
+{\delta_\alpha}^\beta a_\mu -g_{\mu\alpha}a^\beta
-{\varepsilon^\beta}_{\mu\alpha\gamma}b^\gamma)l_\beta
\ee
containing the Weyl nonmetricity 4-vector $a_\mu(x)$ and the skew symmetric
torsion pseudo-trace 4-vector $b_\mu(x)$ represented respectively by the real
and imaginary parts of the unique complex gauge field (GFC)
$\phi_\mu(x)=a_\mu(x)+ib_\mu(x)$.
Effective Weyl-Cartan connections of the type (\ref{realweylc}) were introduced
(for the case of the flat metric background $g_{\mu\nu} = \eta_{\mu\nu}$)
in~\cite{K95,KR2000}, in the framework of {\it noncommutative analisis} over
the algebra of {\it biquaternions}. It has been also applied to formulate a geometrical version of
the Weinberg-Salam electroweak theory~\cite{Krechet}. Finally, it was dealt
with in~\cite{T96} in connection with the theory of Einstein-Weyl spaces
and the ``heterotic geometry'' arising in the SUSY nonlinear $\sigma$-models.

We are now in a position to formulate the main result of the section.

\bt  The following three statements for null SFC on an
algebraically special Riemannian manifold are equivalent:

1.~SFC is a special one (SSFC), i.e. admits a defining equation of the reduced
form (\ref{red}).

2.~SFC defines a repeated principal null direction of the Weyl tensor.

3.~Tangent vector field of a SFC is parallel with respect to effective
Weyl-Cartan connection of the form (\ref{realweylc}).

\et

To conclude, we mention that in the space of Petrov types D (or N) both
SFC (or one single SFC) if exist are of a special form (i.e. SSFC).

\section{Gauge (electromagnetic) fields associated with shear - free null
congruences}

We are going now to study the properties of the GFC for a SSFC on a
curved Riemannian background. By this it is noteworthy that the SSFC defining
system (\ref{red}) possesses a {\it residual} ``weak'' gauge invariance of
the form
\be \l{resgauge}
\xi_A\mapsto \alpha\xi_A, \qquad \phi_\mu \mapsto \phi_\mu + \partial_\mu
\ln \alpha
\ee
where the gauge parameter $\alpha(x)$
can not be an arbitrary function of coordinates. Indeed,
it is easy to check that the gauge parameter should satisfy the equation
\be \l{reseq}
\xi^A\n_{AA'}\al=0
\ee
For analytic congruences there always exists two functionally independent
solutions of this equation \cite{PR86} so that $\alpha$ is constant on the two
complex 2-surfaces canonically defined by SFC. In flat space it means that
$\al$ is a function of the twistor $\mathsf{T}_a=(\xi_A, \xi_A X^{AA'})$
corresponding to a ray of SSFC.

Thus, symmetry (\ref{resgauge}, \ref{reseq}) is a generalization of the ``weak''
gauge symmetry for the SSFC defining equation in the flat case~\cite{K95,KR2000}.
In view of this gauge invariance we
indeed may treat the GFC, i.e. the field represented by the potentials
$\phi_{AA'}$, as an electromagnetic-like field associated with (generated by)
a SSFC.

However, it is necessary then to derive the dynamical equations of the GFC.
For this, consider now the complete set of integrability conditions for the SSFC
equation (\ref{red}) of general type.
Rearranging the covariant derivatives and splitting the arising curvature
spinor into canonical irreducible parts $\Psi_{(ABCD)}$,
$\Phi_{(A'B')(CD)}$ and $\Lambda$ we obtain
\bg
\Psi_{ABCD}\xi^D=\xi_{(A}\varphi_{BC)} \l{IC1}\\
6\Lambda\xi_A=\varphi_{AB}\xi^B-\frac{3}{2}\xi_A\Pi \l{IC2}\\
\Phi_{(A'B')(CD)}\xi^D=\pi_{(A'B')(CD)}\xi^D+\frac{1}{2}\t\varphi_{A'B'}\xi_C
\end{gather}
where the spinors $\varphi_{AB}:=\nabla_{A'(A}\phi_{B)}^{A'}$ and
$\t\varphi_{A'B'}:=\nabla_{A(A'}\phi_{B')}^{A}$ are respectively the antiself-
and selfdual parts of the complex strength tensor of the GFC
\be
{\eu F}_{\mu\nu}:=\nabla_{[\mu}\phi_{\nu]}
=\varphi_{AB}\varepsilon_{A'B'}+\t\varphi_{A'B'}\varepsilon_{AB}
\ee
and where also $\Pi:=\nabla_{\mu}\phi^{\mu}+2\Phi$,
$\pi_{(A'B')}^{(CD)}:=\nabla_{(A'}^{(C}\phi_{B')}^{D)}-\phi_{(A'}^{(C}\phi_{B')}^{D)}$
and $\Phi:=\phi_\mu\phi^\mu=\frac{1}{2}\phi_{AA'}\phi^{AA'}$.

In the conformally flat space or in a space of the Petrov type N the left side
of (\ref{IC1}) vanishes, $\varphi_{AB}=0$, and the GFC ${\eu F}_{\mu\nu}$ is
necessarily selfdual. Taking then into account the existence of 4-potential
$\phi_\mu$ (the exactness of the 2-form ${\eu F}$), we obtain that the complex
GFC ${\eu F}_{\mu\nu}$ (as well as its real part) satisfies homogeneous
Maxwell equations.

In the case of space of the Petrov type III with the SSFC defining
the triple PND one has $\Psi_{ABCD}\xi^C\xi^D=0$. Then we get from
Eq.(\ref{IC1}) $\varphi_{AB}\xi^B=0$ so that $\varphi_{AB} =
\lambda \xi_A \xi_B$ with $\lambda(x)$ being a complex function.
The last expression shows that in this case the {\it antiselfdual} part of the
GFC strength has the familiar form of Robinson's null field~\cite{Rob61}.
However, under considered conditions the Robinson's theorem does not hold since
the gauge freedom has been already used to vanish the spinor $\eta^{A'}$,
i.e. to bring the congruence to the reduced form.

Generally, making use of Eq.(\ref{red}) it is easy to check that for any SSFC
the null {\it Robinson-like} field $\psi_{AB}=\xi_A\xi_B$
satisfies the following equation:
\be
\n^{AA'}\psi_{AB} = -\phi^{AA'}\psi_{AB},
\ee
This equation links together the Robinson-like field $\psi_{AB}$ and
the considered GFC represented by the 4-potentials $\phi_{AA'}$ and, moreover,
is gauge invariant in the ``weak'' sense, i.e. invariant under the
transformations of the type (\ref{resgauge}, \ref{reseq})~\footnote{Contrary to the GFC,
the field $\varphi_{AB}$ does not remain unchanged in these transformations
being instead rescaled}.

Let us pass now to general case of a SSFC in an algebraically special space
and express the 4-potentials $\phi_{AA'}$ through the spinor field $\xi_A$.
Contracting Eq.(\ref{red}) with an arbitrary independent spinor $\tau^A$ and
introducing the normalized spinor $\iota^A:=(\xi_B\tau^B)^{-1}\tau^A$ such that
$\xi_A\iota^A=1$ we obtain for the potentials
\be\l{poten}
\phi_{AA'}=\iota^C\n_{CA'}\xi_A .
\ee

We recall here that in the flat space the GFC (\ref{poten}) not only
satisfies the homogeneous Maxwell equations but also possesses the charge
quantization property. Specifically, the effective electric charge for
every bounded singularity of this field (calculated via making use of the
{\it Gauss theorem}) is self-quantized, i.e. discrete and integer
multiple to a minimal charge (equal in dimensionless units to $\pm 1/4$)
which can be choosed as an analogue of the elementary one~\cite{prot,K2003}.
The property is therein a consequence of the over-determined structure of SFC
equations or, equivalently, of the topological restrictions resembling those
responsible for quantization of the Dirac magnetic monopole~\cite{Dirac}.

Generally, in an algebraically special space the GFC $\varphi_{A'B'}$ is not
necessarily selfdual and, generically, does not satisfies the homogeneous
Maxwell equations. However, if the space is {\it asymptotically flat}, in view of
Eq.(\ref{IC1}) it satisfies the Maxwell equations with {\it effective sources} of
geometrical origin defined by the (derivatives of) conformal curvature of the manifold.
Correspondent corrections to electromagnetic field caused by
the presence of these (extended) sources are proportional to $1/r^3$ and
considerable only at distances of the order of gravitational radius of field
distribution. As to the effective electric charge which is calculated at the
asymptotic, {\it it remains self-quantized and, as before, multiple in value
to the minimal elementary one}.

As a simple example, let us consider one of the radial SSFC, say
$l=(dt-dr)/(1-\cos{\theta})$ of the Schwarzschild space-time. Calculating
the (real part of) associated GFC we find that it is electric in nature, with
a single nonzero radial component $E_r = q(1/r^2 -6Gm/(c^2 r^3))$
where the dimensionless charge $q=1/4$ is equal to elementary $q=e$ in physical
units. The second term is proportional to $1/r^3$, produced by the volume
charge density $\rho = 3eGm/(c^2 r^4) = 3eR_{grav}/(2 r^4)$ and comparable with
the first Coulomb term at a distance $r\approx R_{grav}$.

\section{Invariant spinor differential operators}

Let us derive now some additional fundamental constraints which hold for the principal
spinor of a SSFC. In fact, for SFC in Minkowski space it was proved in~\cite{KW79}  (see
also~\cite{KR2000}) that every component of SSFC spinor $\xi_A$ satisfies the eikonal
equation flat space while their ratio satisfies the wave equation.

We are going now to generalize these equations to the case of a SSFC in on
arbitrary algebraically special space. For this, taking into account the
spinor (i.e., not scalar) nature of the field functions,  we must firstly
write correspondent equations in a manifestly invariant form. In result we come
to the following statement.

\bt For any SSFC the spinor $\xi_A$ satisfies the two sets of equations of the
form
\bea \l{d1}
~E_{(BC)}(\xi)\equiv \nabla_{AA'}\xi_{(B}\nabla^{AA'}\xi_{C)}=0\\
\l{d2}
D(\xi)\equiv \xi^C\nabla^{AA'}\nabla_{AA'}\xi_C=0
\eea
\et

\noindent
{\bf Proof.} To check the first three equations  it's sufficient to use
expressions for covariant derivatives of $\xi$ from the SSFC equation (\ref{red}).
On the other hand, differentiating Eq.(\ref{red}) and taking then the
contraction with $\xi^B$ one gets $D(\xi)=\varphi_{AB}\xi^A\xi^B$, and
the last term vanishes for every SSFC in consequence of the integrability
condition (\ref{IC2}).

\bigskip

As the defining Eq.(\ref{red}) of SSFC, so the Eqs.(\ref{d1},\ref{d2}) are
invariant under ``weak'' rescalings (\ref{resgauge},\ref{reseq}) of the spinor
$\xi$. This means, in particular,
that essential are only the restrictions they impose on the {\it ratio} of
two components of the spinor $\xi_A$, say, on the function $G=\xi_1/\xi_0$.
In flat case from Eqs.(\ref{d1},\ref{d2}) one gets two equations for
$G(x)$, specifically the eikonal equation $\eta^{\mu\nu} \partial_\mu G
\partial_\nu G = 0$ and the linear wave equation $\eta^{\mu\nu}
\partial_\mu\partial_\nu G = 0$.

Thus, the above introduced invariant differential operators $E_{(BC)}(\xi)$ and $D(\xi)$
can, in fact, be regarded as {\it spinor analogues of the two known Beltrami operators}
the latter acting on scalars. On the other hand, these operators both nullify the
principal spinor of every SSFC.

\end{document}